# High quality and wafer-scale cubic silicon carbide single crystals

Guobin Wang[1,2]†, Da Sheng[1,2]†, Yunfan Yang[1,2], Hui Li[1,2]*, Congcong Chai[1,2], Zhenkai Xie[1,2], Wenjun Wang[1,2], Jian-gang Guo[1], Xiaolong Chen[1,2]*

[1]Beijing National Laboratory for Condensed Matter Physics, Institute of Physics, Chinese Academy of Sciences, Beijing, 100190, China

[2]University of Chinese Academy of Sciences, Beijing 100049, China

*Corresponding authors: lihui2021@iphy.ac.cn; chenx29@iphy.ac.cn

†These authors contributed equally to this work.

**Silicon carbide (SiC) is an important semiconductor material for fabricating power electronic devices that exhibit higher switch frequency, lower energy loss and substantial reduction both in size and weight in comparison with its Si-based counterparts[1-4]. Currently, most devices, such as metal-oxide-semiconductor field effect transistors, which are core devices used in electric vehicles, photovoltaic industry and other applications, are fabricated on a hexagonal polytype 4H-SiC because of its commercial availability[5]. Cubic silicon carbide (3C-SiC), the only cubic polytype, has a moderate band gap of 2.36 eV at room-temperature, but a superior mobility and thermal conduction than 4H-SiC[4,6-11]. Moreover, the much lower concentration of interfacial traps between insulating oxide gate and 3C-SiC helps fabricate reliable and long-life devices[7-10,12-14]. The growth of 3C-SiC crystals, however, has remained a challenge up to now despite of decades-long efforts by researchers because of its easy transformation into other polytypes during growth[15-19], limiting the 3C-SiC based devices. Here, we report that 3C-SiC can be made thermodynamically favored from nucleation to growth on a 4H-SiC substrate by top-seeded solution growth technique (TSSG), beyond what's expected by classic nucleation theory. This enables the steady growth of quality and large sized 3C-SiC crystals (2~4-inch in diameter and 4.0~10.0 mm in thickness) sustainable. Our findings broaden the mechanism of hetero-seed crystal growth and provide a feasible route to mass production of 3C-SiC crystals, offering new opportunities to develop power electronic devices potentially with better performances than those based on 4H-SiC.**

The physical-vapor-transport (PVT) method is the state-of-the-art technique for growing 4H- and 6H-SiC crystals. This involves heating raw SiC powder above 2000 °C to produce gas species containing Si and C, which are then transported to a cold end where crystallization occurs on a seed[20,21]. This process, however, does not work well when it comes to grow 3C-SiC as a higher Si/C ratio in gas species is required. A modified method, close-space PVT, by which a high enough Si/C ratio can be created by separation of raw SiC powder and seed 1~2 mm, allows to grow 3C-SiC[22,23]. This is not a practical pathway to mass production considering the very limited thickness. In addition, early attempts to grow 3C-SiC from high temperature melts are not successful either on 6H- or 4H-SiC seeds because these two polytype inclusions always coexist along the grown 3C-SiC[18]. Alternatively, 3C-SiC films are directly deposited on Si substrate then further process into devices on it. But the large lattice mismatch (~19%) and thermal expansion mismatch (~8%) result in too high density of defects, significantly deteriorating the performances of devices[7]. Recently, a reduction in defects for 3C-crystals can be achieved by further PVT on a free-standing single crystal first prepared by chemical vapor deposition on Si substrates[24]. But the grown boules' thickness and the efficiency are still problematic towards mass production of wafers although *in-situ* switch between the two involved growth methods is feasible[25].

Structurally, 3C-SiC differs from 4H-SiC in the stacking of identical Si-C bilayers[26]. In 3C-SiC, the bilayers are stacked as a crystallographic plane (111) in the sequence ABC[26]. In contrast, in 4H-SiC, the bilayers are stacked as a (0001) plane in the sequence ABCB[26]. The two stacking ways do not cause a significant difference in formation energy, typically a few meV per formula higher for 3C-SiC than for 4H-SiC at zero Kelvin[26]. At temperatures around 1727 ℃, the energy difference between 3C-SiC and 4H-SiC widens to about 5~10 meV per formula, enhancing the stability of 4H-SiC further. However, it is not clear why 3C-SiC is often found as inclusions in 4H-SiC films deposited at around 1650 ℃. Ramakers *et al.*[27] proposed that surface tension plays a crucial role in stabilizing 3C-SiC over 4H-SiC and 6H-SiC, as the former has surface tension that is 20~150 meV per SiC lower than the latter two. This also means that the 3C polytype may be energetically favored over a certain temperature range if surface tension contributes significantly to the change in the overall formation energy, which depends on different surface reconstruction configurations.

**Considerations on stabilizing 3C-SiC over 4H-SiC and crystal growth**

We start off our exploration of growing 3C-SiC single crystals by employing the TSSG. Our strategy is based on two primary considerations. First, the interfacial energy between SiC and melts can be more easily adjusted through altering their chemical compositions in TSSG in comparison to PVT, in which only interface between SiC and gaseous phase exists. Liquid phases are generally thought to be has a more significant effect in changing the interfacial energy than gaseous phases do. It is possible to achieve a lower enough interfacial energy for 3C polytype than for 4H, which will prioritize the nucleation and subsequent growth for former, and suppress that for the latter. Second, 4H-SiC crystals larger than 4-inch can be successfully obtained by TSSG at 1700~1800 °C[28]. In this work, we demonstrate that our strategy works well and bulky 3C-SiC crystals up to 4-inch in diameter and more than 4.0 mm in thickness are successfully grown.

Fig. 1a shows the schematic setup for growing 3C-SiC by TSSG. Crucibles made from high purity graphite serve as container and carbon source. Inside the crucible, a temperature gradient is set as 5~15 °C/cm with a temperature of top melt at about 1850 °C by induction heating. The melt is usually composed of Cr, Ce and Si, which become a liquid above 1680 °C (Supplementary Fig. 1) and act a flux having a solubility of C depending on temperature and composition. Three basic steps are involved in the growth process. 1) the flux dissolves the crucible bottom and 10~15 at. % C enter into the melt[28]. 2) thermal conventions convey these C atoms from the bottom to top and 3) the C and Si atoms will combine and crystallize onto the seed as SiC crystal where the temperature is several to a dozen of degrees lower, see Fig. 1b. The stable growth of SiC crystal requires the C flow is at equilibrium among these three steps. In a typical run, we use commercial semi-insulating 4H-SiC (0001) wafers as seed and the growth is performed under a mixed Ar/N$_2$ gas.

For a typical vicinal (0001) surface, the Gibbs free energy change $\Delta G_{homo} = \pi r^2 h \cdot \Delta g + 2\pi r h \cdot \sigma_{4Hside}$, if a two-dimensional 4H-SiC nucleus with a radius of $r$ forms on a 4H-SiC step terrace. In comparison, the change $\Delta G_{hetero} = \pi r^2 h \cdot \Delta g + 2\pi r h \cdot \sigma_{3Cside} + \pi r^2 \cdot (\sigma_{3C/melt} - \sigma_{4H/melt}) + \pi r^2 \cdot \sigma_{3C/4H}$, if a two-dimensional 3C-SiC nucleus on a 4H-SiC step terrace, where $\Delta g$ is the Gibbs free energy change from liquid to solid per volume, $\sigma_{4Hside}$, $\sigma_{3Cside}$, $\sigma_{4H/melt}$, $\sigma_{3C/melt}$ the interfacial energies between lateral surfaces, (0001), (111) facets to melts for 4H- and 3C-SiC, respectively; $\sigma_{3C/4H}$ the interfacial energy for (0001) and (111) crystallographic planes between the two polytypes, $h$ the height of the island. It is reasonable to assume that $\sigma_{4Hside} \approx \sigma_{3Cside}$ because these lateral surfaces form from stacking Si-C bilayers in a similar spacing but in a different sequence, their interfacial energies will approach equal if averaging the fluctuations of interactions at a macro-scale. $\sigma_{3C/4H} \approx 0$ is a reasonable assumption because of the negligible lattice mismatch between 4H-(0001) and 3C-(111). Therefore, the $\Delta G_{hetero}$ is always smaller than the $\Delta G_{homo}$ if the $\sigma_{3C/melt} - \sigma_{4H/melt} < 0$. This means that nucleation and crystal growth are favored for 3C than for 4H if the difference between $\Delta G_{homo} - \Delta G_{hetero}$ is large enough. It is expected that nucleation easily occurs on 4H substrate and its step flow is faster than that for 4H, leading to the total coverage of 3C-SiC on 4H-SiC substrate. Then the growth of 3C will proceed steadily. Fig. 1c schematically describes the possible route for the phase transition starting from preferential hetero-nucleation to subsequent growth for 3C-SiC single crystal on the condition that it has a lower enough interfacial energy with melts. In this study, it is found that the $\sigma_{3C/melt} - \sigma_{4H/melt}$ is negative enough when $N_2$ partial pressures ($p_{N_2}$) above the melts is in the range of 15~20 kPa, justifying the above arguments and expectations. Fig. 1d-f and Supplementary Fig. 2 a, b show the photographs for 2~4-inch 3C-SiC crystal boules grown under $p_{N_2}$ of 20 kPa, respectively. The thickness varies between 4.0~10.0 mm in an 84 h-long growth duration (Table 1). The growth rate is about 50~113 μm/h, a little bit lower than 150 μm/h for the PVT method[20]. The 1 mm thick wafers are black in color (Supplementary Fig. 2c) because of high carrier density introduced by N-doping. It is green color under strong light (Fig. 1g)[29].

**Structural, defective and property characterizations**

Raman scattering measurement are performed on total 20 sites across the entire wafer surface. All spectra are nearly same, only the peaks at 796 cm[-1] are present (Fig. 2a, b). The peak is assigned to be the 3C-SiC's transversal optical mode (TO)[30-32]. Another characteristic mode, the longitudinal optical one, located at 975 cm[-1], which is dependent on the carrier density, does not appear. No folded transverse optical modes at 776 and 707 cm[-1] for 4H- and 6H-SiC are observed[32,33]. A small peak (marked by an arrow) at 741 cm[-1] are probably due to the stacking faults or stress[30,31]. To obtain the information on the evolution of the transition from 4H to 3C, we did the Raman scattering on a cross section of the grown boule and the results are shown in Fig. 2b and Supplementary Fig. 3. It is clearly seen that the 3C occurs immediately at the upper surface of the seed, following a transition zone (TZ) about 20 μm consisting of both 3C and 4H. Then the single phase of 3C is maintained throughout the boule. The observation of photoluminescence (PL) at 523 nm (Fig. 2c) corresponds well to the bandgap of 2.36 eV, confirming the 3C polytype.

The boule surface is quite flat, but growth steps from 9~22 nm are clearly seen (Supplementary Fig. 4). Single crystal and powder X-ray diffraction on small grains cracked from the boule confirms the polytype is 3C with refined lattice parameter $a = 4.3563(4)$ Å (Supplementary Fig. 5, Supplementary Tables 1 and 2), similar to the previously reported results[7]. Plan-view high-angle annular dark field scanning transmission electron microscopy (HAADF-STEM) taken on a spherical aberration TEM (Fig. 2d) clearly identify Si and C atoms arrayed in a manner of ABC sequence and the selected area diffraction pattern in inset of Fig. 2d along $[1\bar{1}0]$ zone axis (Z.A.) can be indexed based on a space group of $F$-43m. Electron energy loss spectrum (EELS) mapping results (Supplementary Fig. 6) indicate the homogeneous distribution of C and Si at a nanoscale level. Energy dispersive spectroscopy (EDS) mapping results also indicate the homogeneous distribution of Si, C and N (Supplementary Fig. 7).

The crystal grows by stacking of (111) crystallographic planes as only the diffraction peaks (111) and (222) are present in the $\theta$-$2\theta$ scan on the surface of the grown boule, see Fig. 3a. To assess the crystallinity of the wafer, we perform the X-ray rocking curve (XRC) measurement. The full width at half maximum (FWHM) for as-grown (111) surface (Fig. 3b) ranges from 28.8 to 32.4 arcsec with an average of 30.0 arcsec (Table 1). The FWHM is very homogeneous across the whole wafer, indicating the high uniformity. To our best knowledge, this value stands for the best results on wafers larger than 2-inch obtained so far (Supplementary Table 3). Defects are characterized on the wafer after being etched at 500 °C for 10 min in KOH melt. Linear ridges, triangle pits and rounded-triangle pits are clearly seen on Si-terminated surface under an optical microscope (OM) and a scanning electron microscope (SEM) (Fig. 3, c-f). The ridges from dozens to more than one hundred microns in length are due to the stacking fault (SF) (Fig. 3, c and f, Supplementary Fig. 8), a common defect in 3C-SiC[8,34-36]. The thickness of typical SFs revealed by the bright-field and dark-field TEM images (Fig. 3g, h) is 3 layers of (111) planes. Its density, defined by the total length of all SFs divided by the observed area, is averaged to be 92.2 /cm (Table 1, Supplementary Fig. 9), much less than what's previously reported (Supplementary Table 4). Calabretta $et~al.$[37] found that N-doping can substantially increase the SFs length, in good agreement with our observations here. In addition, the SFs, seen from a slice of 3C-SiC, are delimited by two triangle pits or two rounded-triangle pits (Supplementary Fig. 10). The typical triangle pits are ~5 μm in size, probably originating from thread screw dislocations (TSDs) (Fig. 3c, d). The rounded-triangle pits, a little bit smaller in size, are from thread edge dislocations (TEDs) (Fig. 3c, e, Supplementary Fig. 9)[35,36,38-40]. They are about $4.3 \times 10^4$ /cm² and $13.9 \times 10^4$ /cm² in density, respectively (Supplementary Fig. 9). No double-positioning boundaries (DPBs), which are quite common in 3C-SiC[22,25,41-43], are observed in our 3C-SiC wafers.

The electrical characterizations are conducted on a slab crystal cut from the grown boules. Electrical resistivity, carrier density and mobility are measured by the standard six-wire method (Supplementary Fig. 11). Supplementary Fig. 12a shows the variations of electrical resistivity with temperature from 5 to 300 K. We can see that the samples grown under $p_{N_2}$ of 15 and 20 kPa exhibit a metallic character. The resistivity decreases with lowering temperature, suggesting the 3C-SiC should become a semi-metal with a room-temperature resistivity 0.58 m$\Omega$·cm (Table 1, Supplementary Table 5 and 6), more than one-order lower than the 4H-SiC's (15~28 m$\Omega$·cm) (Supplementary Table 8)[44]. We note that the crystal grown with $p_{N_2}$ of 10 kPa behaves like a semiconductor below about 100 K (Supplementary Fig. 12a). The carrier density for the 20 kPa sample is calculated to be $1.89 \times 10^{20}$ /cm$^3$ (Supplementary Fig. 12b and Supplementary Table 5), in good agreement with the measured N concentration $1.99 \times 10^{20}$ /cm$^3$ by SIMS (Supplementary Fig. 13). It demonstrates that almost all of the doped electrons are activated to the conduction band at room temperature. The calculated mobility ranges from 56.95 to 62.66 cm$^2$/V s (Supplementary Table 5). The mobility can be enhanced to be 66.24 cm$^2$/V s when the carrier density is lowered (Supplementary Fig. 12c, Supplementary Tables 5 to 7), meanwhile the resistivity mounts up to 5.77 m$\Omega$·cm, about one fourth of 4H-SiC's (15~28 m$\Omega$·cm) at room temperature[44], which is much lower than the reported results (Supplementary Table 8). In this case, the PL at about 523 nm due to the band-edges transition is observed, as state above, see Fig. 2c.

**Measurements of surface tension and contact angles of melts to substrates**

To justify our arguments that the interfacial energy plays an important role in growing 3C-SiC crystal, we measured the surface tension of melts and their contact angles with 3C- and 4H-SiC crystals at high temperatures and different $p_{N_2}$. Fig. 4a, b are the photographs of the liquid drop at 1850 °C under $p_{N_2}$ of 20 kPa (Melt 4) on 3C-(111) and semi-insulating 4H-SiC (0001) crystals, respectively, indicating the contact angles are 40.38 °±0.64 °and 45.55 °±0.07 °. The measured surface tension is $\sigma_{Melt\ 4}$ = 761.24 ±27.83 mN/m at the same temperature (Fig. 4c and Supplementary Figs. 14 and 15). According to Young's equation $\sigma_{SiC/Melt\ 4} = \sigma_{SiC} - \sigma_{Melt\ 4} \cos\theta$, it is easily obtained that the interfacial energies for 3C and 4H are $\sigma_{3C/Melt\ 4}$ = 2151.11 ±21.90 mN/m and $\sigma_{4H/Melt\ 4}$ = 2390.89 ±19.50 mN/m (Supplementary Table 9), where $\sigma_{SiC}$ is estimated from the results in Ref. 27. Fig. 4d shows the variations of melt surface tension, the contact angles of melt on substrates and the calculated interfacial energy between the melt and 3C-(111) and 4H-(0001) of single crystals based on Young's equation with $p_{N_2}$ measured for many times (Supplementary Figs. 14 and 15). The melt's surface tension decreases as increasing $p_{N_2}$, which can be attributed to the dissolved N in the melt (Supplementary Fig. 16). We can see that interfacial energy between the melt and 3C-(111) is lower than that for 4H- (0001), and their difference widens with the increasing $p_{N_2}$. Our results indicate that other polytype inclusions are present when the $p_{N_2}$ is below 10 kPa. Optimal pressures of $p_{N_2}$ ranging from 15 to 20 kPa are required to stabilize the 3C polytype during the growth. We then regrow the crystal on a 3C-SiC seed using the same compositions of flux under 20 kPa $N_2$ again. Raman scattering measurements confirm the 3C polytype (Supplementary Fig. 17).

The results presented here demonstrate that bulk 3C-SiC crystals can be grown through changing the interfacial energy of melt. More importantly, this TSSG route provides a reliable method to grow high-quality wafer-scale 3C-SiC, exhibiting the potential for further mass production. Thus-grown crystals are very suitable used as substrates for homogeneous epitaxy and device fabrication in terms of high-crystallinity, high conductivity and availability. Better homoepitaxy 3C films and the power devices are expected to be fabricated, and thus boost the SiC industry further. Alteration of interfacial energy reported here could be applied to other layered materials to obtain the single crystals that otherwise are difficulty to grow.

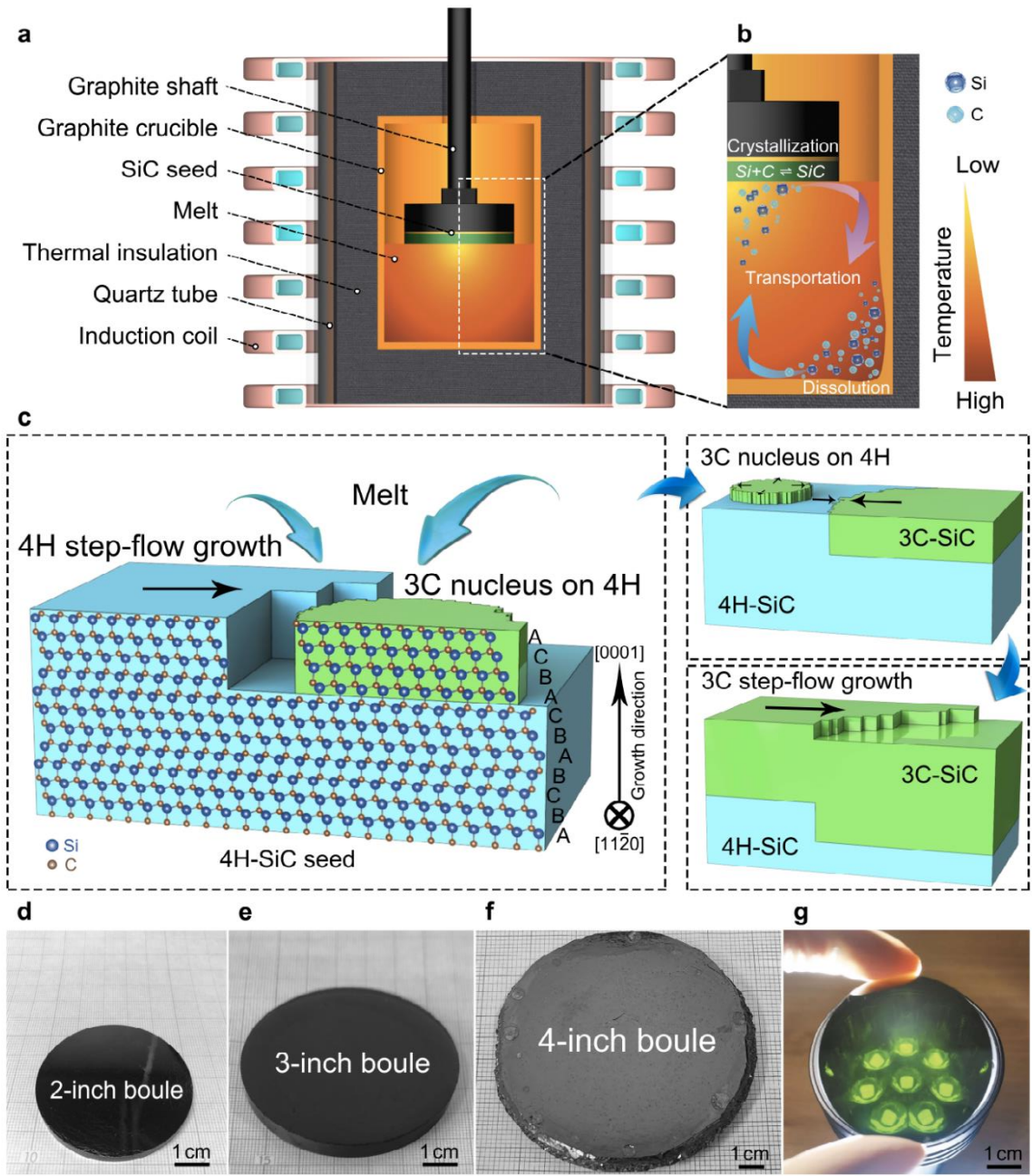

**Fig. 1 | TSSG growth of 3C-SiC single crystals. a**, Schematic of the setup for growing 3C-SiC by TSSG. **b**, Schematic of three basic growth processes for TSSG: 1. Dissolving C from the graphite crucible at high temperature region, 2. Transportation of C from the high temperature region to the low temperature driven by the convection, 3. Crystallization of SiC on the low temperature seed crystal. **c**, Proposed growth model of 3C-SiC on a 4H-SiC seed via TSSG. **d-f**, Photographs of 2-, 3-inch 3C-SiC boule after rounded cutting process and as-grown 4-inch 3C-SiC boule. The thickness of the 2~4-inch 3C-SiC boule is above 4.0 mm. **g**, Photograph of 3C-SiC single crystal wafer.

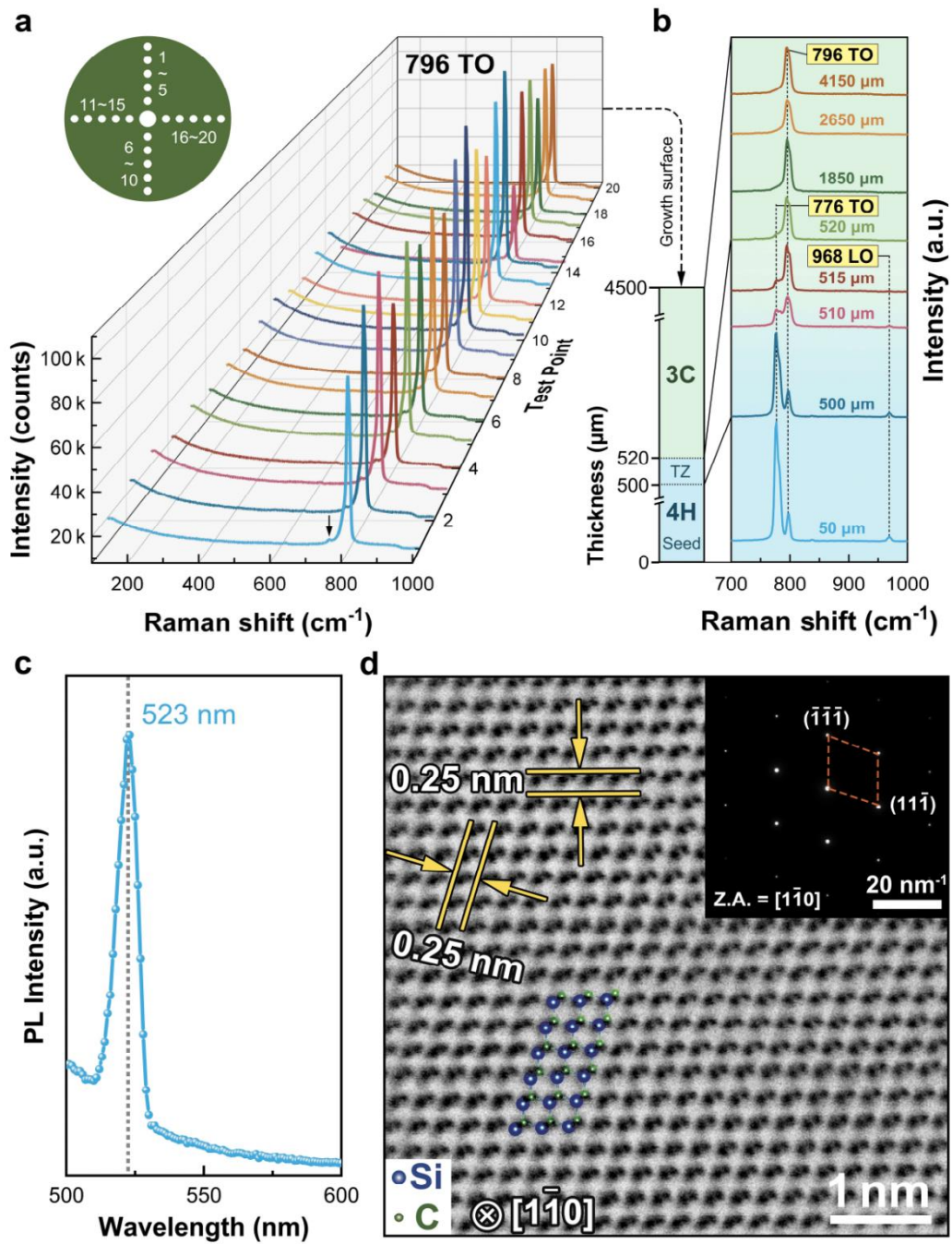

**Fig. 2 | Identification and confirmation of 3C-SiC polytype for as-grown crystals. a**, Raman spectra of 3C-SiC measured on 20 points on the 2-inch crystal. The inset shows the distribution of all measured points. **b**, Raman spectra of seed 4H-SiC, TZ (transition zone) and grown 3C-SiC. **c**, PL spectrum of 3C-SiC measured at 300 K. **d**, Plan-view high-angle annular dark field scanning TEM (HAADF-STEM) image of 3C-SiC. Si and C atoms are superimposed. Inset is SAED measured along $[1\bar{1}0]$ Z.A. (zone axis).

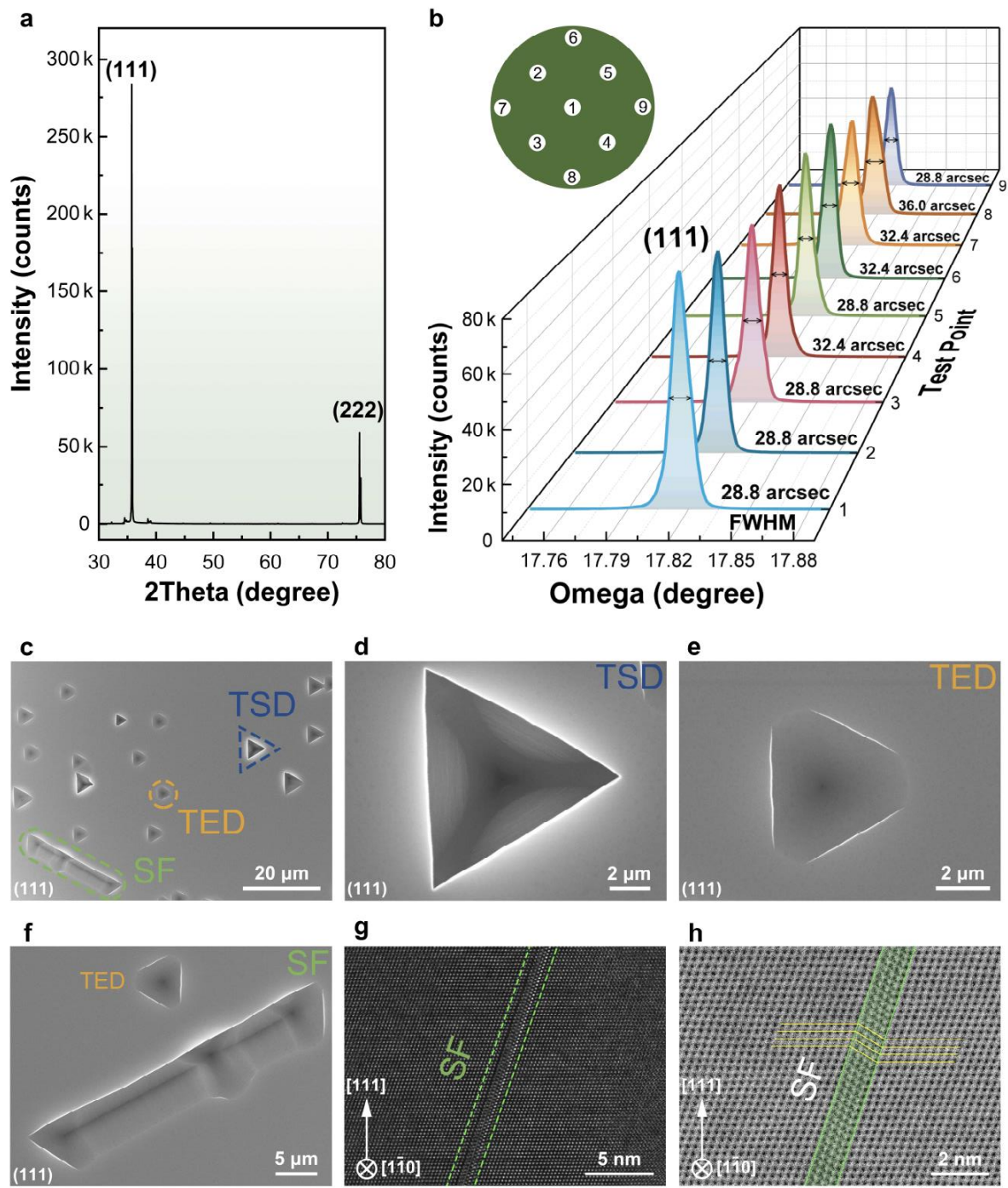

**Fig. 3 | Characterizing the crystallinity and defects of 3C-SiC wafer. a**, X-ray diffraction (XRD) spectrum for 3C-SiC wafer, showing the growing surface is (111) plane. **b**, X-ray rocking curve (XRC) of (111) plane, and the FWHM ranges from 28.8 to 32.4 arcsec. The inset shows the distribution of 9 measured points. **c-f**, OM (Optical microscope) images for 3C-SiC wafer after etch at 500 °C for 10 min in KOH melt, revealing the existence of SF (stacking fault), TSDs (threading screw dislocations), and TEDs (threading edge dislocations) defects in the 3C-SiC wafers. **g**, **h** HAADF-STEM images of a SF composed of three layers of SiC.

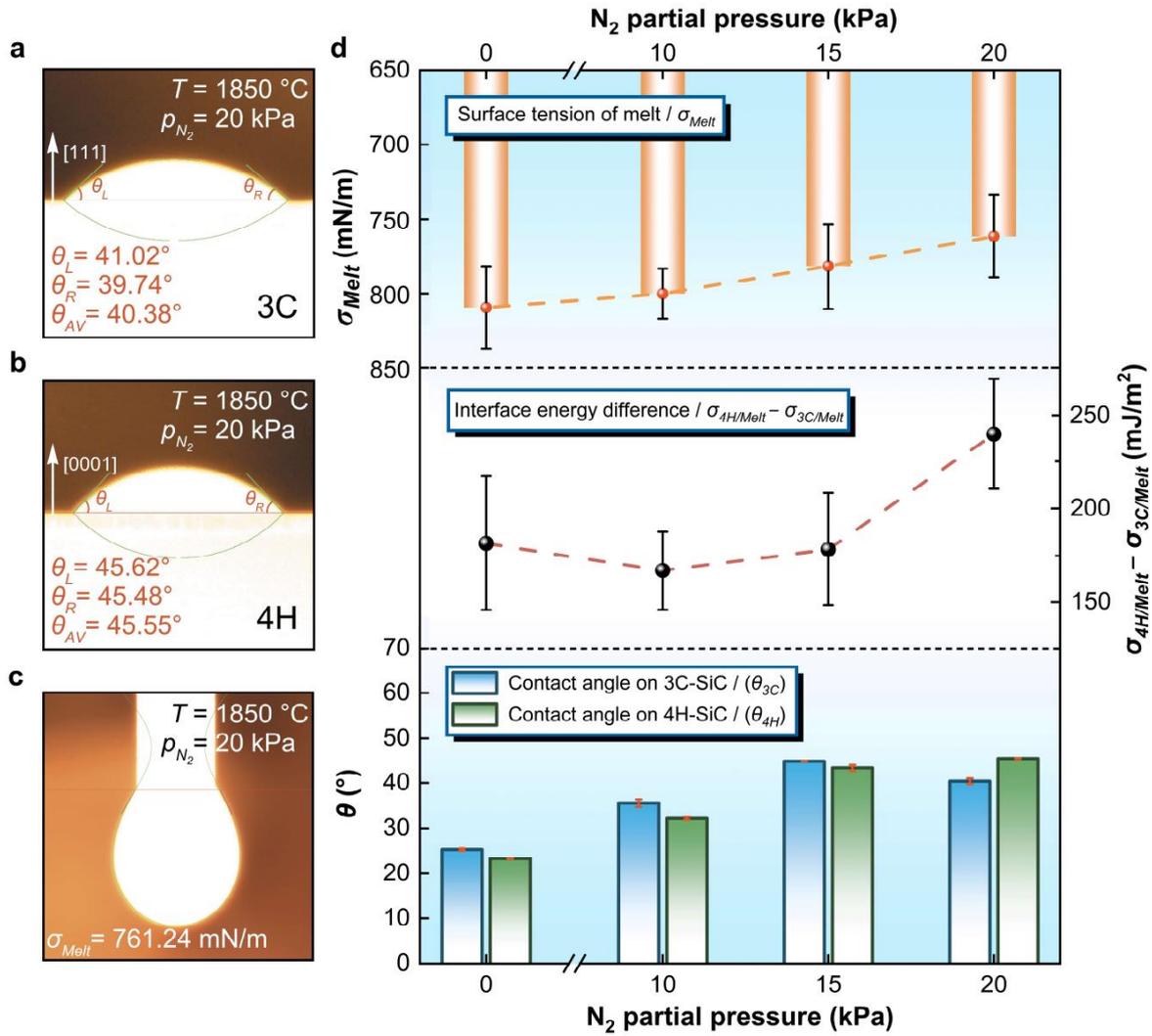

**Fig. 4 | *In-situ* measurements of contact angles and surface tension of melts.** High temperature contact angle of melt droplet of Melt 4 on (**a**) 3C-SiC (111) surface and (**b**) semi-insulated 4H-SiC (0001) surface. The average contact angles are 40.38 ° ± 0.64 ° and 45.55 ° ± 0.07 °, respectively. **c**, Measurement of surface tension of Melt 4 via intravenous drip method. **d**, Histogram of surface tension (upper panel), high temperature contact angles on 3C-SiC (111) and 4H-SiC (0001) planes (lower panel), the solid-liquid interfacial energy difference between melt and 4H-SiC (0001) and 3C-SiC (111) surface (middle panel) of Melt 4. The measurements are carried out at 1850 °C under $p_{N_2}$ of 0 (Melt 1), 10 (Melt 2), 15 (Melt 3) and 20 kPa (Melt 4) with the total growth pressure of 50 kPa.

**Table 1. Properties of 3C-SiC bulk single crystals**.

| Growth Method | Diameter (inch) | Thickness (mm) | FWHM of XRC (arcsec) | SF density (/cm) | Electrical resistivity (mΩ·cm) |
|---|---|---|---|---|---|
| TSSG | 2~4 | 4.0~10.0 | 30.0 | 92.2 | 0.58 |

## Methods

### Growth of 3C-SiC bulk single crystals via TSSG technique

A 4H-SiC seed was stuck onto a graphite seed-supporter with AB glue with Si terminated (0001) surface as growth front. The graphite seed rod was then connected with the graphite seed-supporter. Si lumps of 5-10 cm taken from Si ingot, Cr grains of 3~5 mm in size and bulk Ce rod with a molar ratio of (35~70) : (20~55) : (0.5~10) were put into a graphite crucible. The graphite crucible, used both as the flux container and the carbon source, was put into the growth system. The induction heating furnace was pumped into $10^{-5}$ Pa and then charged into high purity Argon (Ar, 99.9995%, 5N) gas and Nitrogen ($N_2$, 99.9995%, 5N) gas to 50 kPa with the $N_2$ partial pressure of 0 (Melt 1), 10 (Melt 2), 15 (Melt 3) and 20 kPa (Melt 4). After that, the furnace was heated until the temperature of the SiC seed reached to 1700~1900 °C measured by a dual-band infrared thermometer with a resolution of ±0.5%. The temperature gradient from the top to the bottom of the melt with a height of 15~25 mm was 5~15 °C/cm. During the SiC growth process, the seed was rotated at 50~150 rpm while pulled at a rate of 30 μm/h. After growth, the temperature was cooled down to room temperature at a cooling rate of ~100 °C/h. The SiC boule was obtained and characterized.

### SiC single crystal wafers preparation

The as-grown SiC boule was rounded into circled shape with the outer circle diameter of 2~4 inch by the computer numerical control machine cylindrical grinding machine (MK1320, Deyou, China) with a rounded rate of 0.14 mm/min. The rounded SiC boule was then sliced into ~1 mm thickness via single wire cutting machine (CHSX5625-XW, Chenhong, China) or directly sliced into ~650 μm by multi-wire cutting machine (MWS-812SD, Takatori, Japan). The sliced SiC wafers were then ground and polished in turn via single-sided grinding machine (36DPAW-TD) and single-sided polishing machine (36GPAW-TD, Speedfam, China). The polished SiC wafers were characterized and applied to measure the high temperature contact angle with the melts. The polished SiC (111) wafers were also used as the seed to grow 3C-SiC bulk single crystals via the TSSG method.

### Etching of SiC wafers

The as-grown surface of SiC boules and the polished SiC (111) surface were etched in the molten KOH at 500 °C for 10~30 min.

### Characterizations

The as-grown SiC boules and wafers were characterized by X-ray diffraction, OM, SEM, Raman, PL, XPS, TEM, electrical measurements, SIMS and high temperature contact angle and surface tension measurements.

**Data Availability**

All data are available in the main text or supplementary materials. The data and code that support the findings of this study are available from the corresponding authors on reasonable request.

**Acknowledgments**

This work was supported by the Beijing Municipal Science and Technology Project (Grant No. Z211100004821004), the Special Project on Transfer and Conversion of Scientific and Technological Achievements of the Chinese Academy of Sciences (Grant No. KFJ-HGZX-042). We thank YS from Institute of Physics, Chinese Academy of Sciences (IOP, CAS) for the single crystal XRD measurement; SJ from IOP, CAS for the discussion of the single-crystal XRD results; QZ from IOP, CAS for the help of TEM measurements; ZZ, GZ, LY and CP from Beijing Lattice Semiconductor Co., Ltd. for technical help.

**Author contributions**

XC conceived and supervised the project. HL and XC designed this project. GW, DS, HL, YY and WW contributed to the growth of SiC single crystals. DS and GW contributed to the high temperature contact angles and melt surface tension measurements. GW, YY and HL performed the defects characterizations of SiC single crystals, PL, Raman and TEM measurements. HL and CC conducted the single crystal XRD measurements and CC conducted the single crystal XRD refinement. GW and ZX conducted the electrical measurements. GW, DS and HL performed other investigations. XC, HL and JG wrote the manuscript based on discussion of all co-authors. All authors discussed the results and reviewed the manuscript.

**Competing interests**

The authors declare no competing interests.

**Supplementary information**
**Materials and Detail characterization methods**
**Supplementary Figs.** 1 to 17
**Supplementary Table** 1 to 9
**References** 1-42